\documentclass[sigconf]{acmart}
\usepackage{multirow}
\AtBeginDocument{%
  }

\setcopyright{acmlicensed}
\copyrightyear{2018}
\acmYear{2018}
\acmDOI{XXXXXXX.XXXXXXX}
\acmConference[Conference acronym 'XX]{Make sure to enter the correct
  conference title from your rights confirmation email}{June 03--05,
  2018}{Woodstock, NY}
\acmISBN{978-1-4503-XXXX-X/2018/06}

\begin{document}

\title{CogSearch: A Cognitive-Aligned Multi-Agent Framework for Proactive Decision Support in E-Commerce Search}

\author{Zhouwei Zhai}
\authornote{The corresponding author.}
\orcid{1234-5678-9012}
\email{zhaizhouwei1@jd.com}
\affiliation{%
  \institution{JD.com}
  \city{Beijing}
  \country{China}
}

\author{Mengxiang Chen}
\email{chenmengxiang9@jd.com}
\affiliation{%
  \institution{JD.com}
  \city{Beijing}
  \country{China}
}
\author{Haoyun Xia}
\email{xiahaoyun1@jd.com}
\affiliation{%
  \institution{JD.com}
  \city{Beijing}
  \country{China}
}
\author{Jin Li}
\email{lijin.257@jd.com}
\affiliation{%
  \institution{JD.com}
  \city{Beijing}
  \country{China}
}
\author{Renquan Zhou}
\email{zhourenquan.1@jd.com}
\affiliation{%
  \institution{JD.com}
  \city{Beijing}
  \country{China}
}
\author{Min Yang}
\email{yangmin.aurora@jd.com}
\affiliation{%
  \institution{JD.com}
  \city{Beijing}
  \country{China}
}

\renewcommand{\shortauthors}{Zhouwei et al.}

\begin{abstract}
Modern e-commerce search engines, largely rooted in passive retrieval-and-ranking models, frequently fail to support complex decision-making, leaving users overwhelmed by cognitive friction. In this paper, we introduce CogSearch, a novel cognitive-oriented multi-agent framework that reimagines e-commerce search as a proactive decision support system. By synergizing four specialized agents, CogSearch mimics human cognitive workflows: it decomposes intricate user intents, fuses heterogeneous knowledge across internal and external sources, and delivers highly actionable insights. Our offline benchmarks validate CogSearch’s excellence in consultative and complex search scenarios. Extensive online A/B testing on JD.com demonstrates the system's transformative impact: it reduced decision costs by 5\% and achieved a 0.41\% increase in overall UCVR, with a remarkable 30\% surge in conversion for  decision-heavy queries. CogSearch represents a fundamental shift in information retrieval, moving beyond traditional relevance-centric paradigms toward a future of holistic, collaborative decision intelligence.
\end{abstract}




\maketitle

\section{Introduction}
E-commerce search serves as the core hub connecting user needs with product supply. Its fundamental objective extends beyond merely returning relevant products; it aims to assist users throughout the entire cognitive process from “need articulation” to “decision formation.”
Industrial search systems have long evolved around the “retrieval-ranking” paradigm. The retrieval stage has progressed from statistical models like LSI\cite{deerwester1990indexing}, BM25\cite{robertson2009probabilistic}, and TF-IDF\cite{aizawa2003information} to deep semantic matching driven by approaches like RocketQA\cite{qu2020rocketqa}. The ranking stage enhances precision through techniques such as knowledge distillation\cite{liu2022knowledge}, DIN attention mechanisms\cite{zhou2018deep}, and personalized modeling\cite{yan2018beyond}, forming a closed-loop optimization process\cite{subramaniam2025ai}. Recently, Large Language Models have further augmented system performance via query rewriting\cite{dai2024enhancing,peng2024large}, generative retrieval\cite{tay2022transformer,bevilacqua2022autoregressive,li2024generative}, and relevance fine-tuning\cite{thomas2024large,tang2025lref,dong2025taosr1}. However, existing research still faces critical bottlenecks: the passive response mode hinders the system’s ability to handle complex, multi-attribute implicit reasoning needs, and reliance solely on platform-internal data limits cross-platform information integration. More crucially, systems often function as “result displays” rather than “decision facilitators,” leading to excessive cognitive load and comparison costs for users during high-complexity decision-making.
The core challenge in addressing these issues lies in breaking the limitations of static indexing to achieve deep convergence of user intent within dynamic dialogues and transforming fragmented product attributes into structured decision-making bases. Inspired by Simon’s decision-making model\cite{simon1955behavioral}, this paper proposes CogSearch, a cognitively-aligned multi-agent framework. This framework upgrades search from “keyword matching” to “active decision assistance” through the collaborative operation of four specialized agents: Planner, Executor, Guider, and Decider. Our core contributions are:
\begin{itemize}
    \item Architectural Innovation: We propose the CogSearch multi-agent framework. The Planner decomposes complex intents, the Executor integrates multi-source information, the Guider steers demand convergence, and the Decider provides structured purchasing recommendations. This achieves precise alignment between the search process and the user’s shopping cognitive chain.
    \item Large-Scale Industrial Validation: CogSearch has been fully deployed on JD.com, one of China’s largest self-operated e-commerce platforms. Offline experiments demonstrate significant performance gains in handling complex queries. Furthermore, online A/B testing confirms that our framework reduces user decision costs by 5\%, yields a 0.41\% lift in overall User Conversion Rate (UCVR), and achieves a remarkable 30\% UCVR increase for decision-heavy queries.
\end{itemize}

\section{The CogSearch Framework}
\subsection{System Overview}
\begin{figure}[h]
  \centering
  \includegraphics[width=\linewidth]{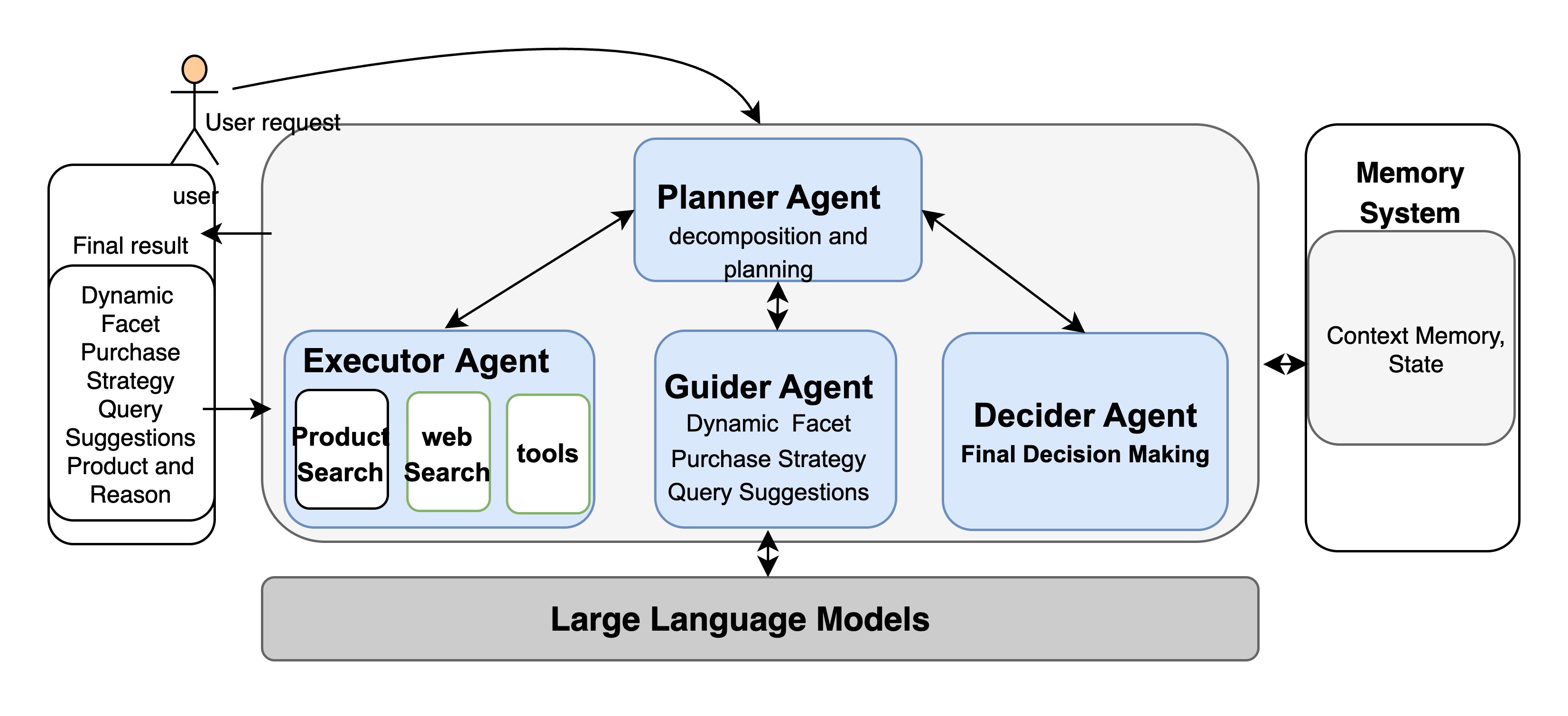}
  \caption{CogSearch Framework Overview}
  \label{figmain}
\end{figure}

Inspired by Simon’s three-stage decision-making model\cite{simon1955behavioral}, CogSearch fundamentally redefines e-commerce search as a collaborative cognitive decision-making process, as opposed to the traditional retrieval-ranking pipeline, as illustrated in Figure ~\ref{figmain}.

The design logic of CogSearch deeply aligns with the user’s shopping cognitive chain:
\begin{itemize}
    \item Planner Agent: Solves the problem of demand understanding. Based on the user’s current query, search history, and historical click context, it comprehends user intent and performs task planning, generating a task graph stored in the Memory System. Specific tasks include product search tasks, web search tasks, and tool/service invocation.
    \item Executor Agent: Solves the problem of multi-source information gathering. Based on the Planning Agent’s output, it executes multi-source retrieval tasks – encompassing in-platform product retrieval (returning a list of eligible products), web-scale information retrieval, and tool/service invocation (e.g., weather, logistics APIs).
    \item Guider Agent: Solves the problem of proactive decision guidance. Leveraging retrieved product results combined with web-sourced information, it generates personalized dynamic filters and purchasing strategies. Furthermore, it proactively identifies latent needs to generate follow-up questions for demand convergence or stimulate new user requirements.
    \item Decider Agent: Solves the problem of integrated decision-making. Synthesizing the user profile, historical behavioral context, interaction signals from the Navigation Agent (e.g., filter clicks), and multi-source information (product attributes + web guides/strategies), it employs LLM-based reasoning to output final product recommendations accompanied by structured decision rationale.
\end{itemize}

Memory System acts as the hub for storing, updating, and maintaining all contextual information for the agents. This includes contextual memories, agent states, and the task graph. 
\subsection{Planner Agent: Intent-Driven Task Generation}

As the cognitive orchestration engine of CogSearch, the Planner Agent transcends traditional query rewriting by formulating the search process as a structured reasoning problem. Its primary objective is to bridge the semantic gap between ambiguous user requirements and precise system execution capabilities through the generation of an executable Task Graph. Formally, we define the planner's input as a dynamic context tuple: 
\begin{equation}
    \mathcal{C}_t = \{q_t, \mathcal{H}^{search}_{<t}, \mathcal{H}^{click}_{<t}\}
\end{equation}
where $q_t$ denotes the current user query, while $\mathcal{H}^{search}_{<t}$ and $\mathcal{H}^{click}_{<t}$ represent the historical query sequence and item interaction context, respectively. The Planner leverages the reasoning capacity of Large Language Models to map $\mathcal{C}_t$ into a Directed Acyclic Graph (DAG), denoted as:
 $\mathcal{G} = (\mathcal{V}, \mathcal{E})$.
The vertex set $\mathcal{V} = \{v_1, v_2, ..., v_n\}$ represents atomic cognitive tasks derived from the user's latent intent. To facilitate comprehensive decision support, we define the action space of $\mathcal{V}$ across three functional dimensions: (1) ProductSearch ($T_{prod}$) for retrieving candidate items from the internal E-commerce inventory; (2) WebSearch ($T_{web}$) for acquiring open-world knowledge (e.g., reviews, usage guides) to support cross-platform information fusion; and (3) ToolInvocation ($T_{tool}$) for executing specific functional operations via the Model Context Protocol (MCP), such as price comparison or logistics querying. The edge set $\mathcal{E}$ explicitly models the semantic dependencies between these tasks. A directed edge $(v_i, v_j) \in \mathcal{E}$ is generated if and only if the execution of task $v_j$ is conditional on the output context of $v_i$, satisfying the constraint $Input(v_j) \subseteq Output(v_i)$.
The generation of $\mathcal{G}$ is modeled as a conditional probability distribution $P_{\theta}(\mathcal{G} | \mathcal{C}_t)$, where the Planner decomposes complex queries into a topological sequence of sub-tasks. 
\subsection{Executor Agent: Multi-Source Acquisition}
\subsubsection{ProductSearch Agent}
This agent transforms structured queries into ranked product candidates via a cognitively augmented hybrid retrieval architecture. To balance precision and diversity: 1) Keyword-Vector Fusion: Concurrently invokes platform APIs (lexical recall) and contrastive learning-based semantic embedding models (intent generalization). 2) Multi-Source Enrichment: Coordinates ProductAttr and ReviewSum sub-agents to aggregate attributes and generate compressed pros/cons summaries from descriptions and reviews.
\subsubsection{WebSearch Agent}
Overcoming platform-bound information gaps, this agent integrates real-time cross-domain data via a noise-robust “Expand-Retrieve-Filter” pipeline: 1)Query Expansion: LLM generates synonymic variants of web search sub-tasks.2) Parallel Retrieval: Fetches raw results from search APIs. 3)Multi-Criteria Filtering: Scores results using:
$$Score(d) = \alpha \cdot Rel(d, q) + \beta \cdot Auth(d) + \gamma \cdot Fresh(d)$$
where $\alpha, \beta, \gamma$ calibrate relevance, authority, and freshness.
\subsubsection{Tool Integration Agent}
Dynamically invokes external APIs (e.g., weather impact on logistics, price trend forecasting) using LLM-generated function calls when contextual triggers are detected in the cognitive loop.

\subsection{Guider Agent: Cognitive-Aligned  Guidance}
\subsubsection{Dynamic Cognitive Facet Generation}
Traditional facets are often pre-defined and rigid. Drawing inspiration from cognitive psychology [23], the Guider Agent employs a Clarify-Guide-Contextualize (CGC) mechanism to generate dynamic, hierarchical filters. By analyzing the latent semantic variance between the retrieved item set $\mathcal{I}$ and the user’s cognitive state $\mathcal{S}_u$, the agent formulates a scoring function to select the most discriminative attributes $a^*$:
$$a^* = \arg\max_{a \in \mathcal{A}} \text{InfoGain}(a | \mathcal{I}, \mathcal{S}_u, \mathcal{C})$$
where $\mathcal{C}$ represents the multi-turn interaction context. This "brain-inspired" hierarchical reasoning [24] ensures that the generated facets (e.g., "Battery Life for Vlogging" vs. "Weight for Hiking") align with the user's real-time decision priorities, significantly reducing the cognitive load of information filtering.
\subsubsection{Personalized Purchase Strategy}
To address the "paradox of choice," the Guider Agent synthesizes multi-source intelligence—including item specifications from the Executor, real-time web knowledge (e.g., expert reviews), and the user’s historical preferences—to generate a Personalized Purchase Strategy.

The strategy is formulated as a structured knowledge prompt:
\begin{itemize}
    \item Expert Anchoring: Identifies key performance indicators (KPIs) for the specific category.
    \item Trade-off Analysis: Explicitly highlights pros and cons based on the user’s "add-to-cart" history and budget constraints.
\end{itemize}
By transforming raw data into actionable "Buying Guides," the agent shifts the system from a passive list provider to a professional shopping consultant.
\subsubsection{Query Suggestions}
The Guider Agent proactively manages the search trajectory by generating dual-purpose suggestions:
\begin{enumerate}
    \item Convergent Queries: Aimed at "zooming in" on specific attributes when the user shows hesitation (e.g., "Noise-cancelling headphones under 200g").
    \item Stimulative Queries: Designed to "expand" the search space based on latent needs identified via web-enhanced cross-platform signals (e.g., "Compatible stabilizers for your new mirrorless camera").
\end{enumerate}
This dual-steering approach ensures that the search process is not merely a reactive loop but a guided journey toward an optimal decision.

\subsection{Decider Agent: Integrative Reasoning and Decision-Making}
The Decider Agent serves as the terminal cognitive nexus of the CogSearch framework. Unlike traditional ranking modules that output probabilistic scores, this agent acts as a reasoning engine designed to reach "cognitive closure." Grounded in Simon’s three-stage decision model (Intelligence-Design-Choice)\cite{simon1955behavioral} and Multi-Criteria Decision Making theory\cite{keeney1993decisions}, the Decider Agent synthesizes heterogeneous signals into actionable, interpretative purchasing advice.

Context Aggregation. The agent first constructs a holistic decision context $\mathcal{S}_{ctx}$ by aggregating outputs from upstream agents. Let $\mathcal{P}$ denote the candidate items and summarized reviews from the ProductSearch Agent, $\mathcal{W}$ the real-time external evidence from the WebSearch Agent, and $\mathcal{H}$ the interaction trajectory from the Guider Agent. Combined with the user’s static profile $\mathcal{U}$ and dynamic constraints $\mathcal{C}$ (budget, preferences) defined by the Planner Agent, the unified context is formalized as:
$$\mathcal{S}_{ctx} = \Phi(\mathcal{P}, \mathcal{W}, \mathcal{H}, \mathcal{U}, \mathcal{C})$$
where $\Phi$ represents the context fusion mechanism. This integration ensures the decision is not isolated but deeply rooted in both internal platform data and the open web ecosystem.

Multi-Criteria Decision Execution. We leverage the reasoning capabilities of large language models to perform MCDM. To mitigate the hallucination risks common in direct generation, we inject a structured evaluation protocol $\mathcal{I}_{eval}$ into the prompt. The agent evaluates candidates based on a composite utility vector $\mathbf{v}$ spanning four dimensions:
\begin{enumerate}
    \item Functional Performance: Assessing core specifications and quality sentiment.
    \item Economic Viability: Analyzing price-to-budget alignment.
    \item Reliability: Cross-verifying internal reviews with external web credibility signals.
    \item Constraint Adherence: Checking compliance with Planner-derived hard constraints.
\end{enumerate}

Decision \& Rationale Generation. The final output is not merely a ranked list but a persuasive recommendation pair $(d^*, \mathcal{R})$, where $d^*$ is the optimal item and $\mathcal{R}$ is the natural language rationale explaining why the item fits the user's specific scenario:
$$(d^*, \mathcal{R}) \leftarrow \text{LLM}(\mathcal{S}_{ctx}, \mathcal{I}_{eval})$$

By explicitly modeling the trade-offs between performance, cost, and reliability, the Decider Agent bridges the gap between system retrieval and user decision-making, providing a transparent path from information seeking to transaction.

\section{Experiments}

\subsection{Experimental Setup}
\subsubsection{Dataset}
To bridge the gap in standardized evaluation for cognitive decision-making in e-commerce search, we propose ECCD-Bench. This benchmark is curated from anonymized JD.com query logs and features 10k expert-annotated query-response pairs. It spans three task categories designed to assess diverse cognitive abilities: 
(1) Simple Needs, involving explicit product or category searches; (2) Complex Needs, requiring advanced reasoning, handling of multiple constraints, negation, or semantic ambiguity; and (3) Consultative Queries, focusing on advisory interactions such as product Q\&A, comparative analysis, and general information seeking with latent shopping intent.
\subsubsection{Evaluation Metrics}
The proposed model is evaluated through both offline benchmarks and online A/B testing.

Offline Metrics: (1) Accuracy@K (ACC@K): Measures the capability to retrieve relevant products aligning with user intent within the top K positions. We report ACC@5 to mirror the real-world display. (2) User Demand Satisfaction (UDS): A human-centric metric (scale 1-3) where experts rate system responses across completeness, correctness, and usefulness.

Online Metrics: (1) Decision Cost (DC): The average number of interactions (searches plus clicks) per transaction. A lower DC signifies higher decision-making efficiency.(2) User Conversion Rate (UCVR): UCVR is defined as the ratio of the total number of completed orders to the total number of Unique Visitors (UV) within a specific timeframe. 
\subsection{Offline Experiments}

We conducted offline evaluations using the current JD.com search system (an LLM-enhanced retrieval-ranking model) as the baseline. To validate CogSearch's core agents and modules, ablation experiments were performed on the ECCD-Bench dataset (Table \ref{tab:eccd_bench}). Results show that CogSearch significantly outperforms the online system in ACC@5 and UDS for complex and consultative queries. This suggests that CogSearch effectively decodes implicit user intent and multifaceted constraints. By mimicking professional shopping assistants through dynamic information synthesis and reasoning, CogSearch excels in complex scenarios, bridging the semantic gap that limits traditional paradigms.

\begin{table}[ht]
\centering
\caption{ECCD-Bench Offline Evaluation}
\label{tab:eccd_bench}
\resizebox{\columnwidth}{!}{%
\begin{tabular}{lccc}
\toprule
\multirow{2}{*}{\textbf{Method}} & \textbf{Simple Needs} & \textbf{Complex Needs} & \textbf{Consultative Queries} \\
 & (ACC@5 / UDS) & (ACC@5 / UDS) & (ACC@5 / UDS) \\ \midrule
online system(base) & 0.87 / 2.64 & 0.48 / 1.92 & 0.60 / 1.81 \\
CogSearch (Full) & \textbf{0.87 / 2.70} & \textbf{0.83 / 2.79} & \textbf{0.65 / 2.88} \\ \midrule
\textit{w/o Planner} & 0.86 / 2.65 & 0.52 / 2.05 & 0.61 / 2.42 \\
\textit{w/o WebSearch} & 0.87 / 2.68 & 0.81 / 2.72 & 0.58 / 2.15 \\
\textit{w/o Guider} & 0.87 / 2.64 & 0.80 / 2.23 & 0.65 / 2.10 \\
\textit{w/o Decider} & 0.85 / 2.62 & 0.75 / 2.38 & 0.63 / 2.40 \\
\textit{w/o Memory} & 0.86 / 2.61 & 0.68 / 2.35 & 0.62 / 2.45 \\ \bottomrule
\end{tabular}%
}
\end{table}
Ablation study findings:
\begin{itemize}
    \item \textbf{w/o Planner:} ACC@5 and UDS for complex queries plummet by 37.3\% and 26.5\%, respectively, while simple queries remain stable. This confirms the Planner's necessity in bridging semantic gaps via intent decomposition.
    
    \item \textbf{w/o WebSearch:} UDS for consultative queries drops by 25.3\%, highlighting the importance of web-scale information fusion for advisory decision-making.
    
    \item \textbf{w/o Guider:} Complex and consultative UDS fall by 20\% and 27\%. Replacing the Guider with static filters increases user cognitive load, demonstrating the value of dynamic, personalized guidance.
    
    \item \textbf{w/o Decider:} A universal decline in UDS underscores that multi-criteria reasoning and rationales are essential for a robust decision loop, transcending simple list re-ranking.
    
    \item \textbf{w/o Memory:} All metrics decrease, validating the memory system’s role in ensuring contextual coherence and persistent intent alignment.
\end{itemize}

\subsection{Deployment and Online A/B Testing}

We integrated CogSearch into the JD.com search platform via an opt-in "AI Search" entry. To satisfy strict online latency requirements, the Planner and Guider modules utilize Qwen3-4B\cite{yang2025qwen3}, while the Decider employs Qwen3-8B\cite{yang2025qwen3}. These models were fine-tuned via knowledge distillation from DeepSeek-R1\cite{guo2025deepseek}. We optimized online inference using INT8 quantization, caching mechanisms, and parallel scheduling. Although the average end-to-end latency is 1.3s (compared to 800ms for traditional search), we implemented streaming output to maintain a responsive user experience.

We conducted a two-week online A/B test with 10\% of live traffic. The experimental group was granted access to the CogSearch interface, while the control group was restricted to traditional search. Results show a significant 5\% reduction in Decision Cost (DC) and an average 0.41\% increase in User Conversion Rate (UCVR) for the experimental group. Notably, for decision-heavy intents—such as professional product interpretation, gift selection, and functional comparisons—UCVR improved by over 30\%.

Behavioral analysis reveals that users leverage CogSearch for more complex, personalized needs; the proportion of long queries (>10 words) increased by 25\% relative to the control, reflecting a shift toward natural language expression. These results demonstrate that CogSearch effectively reduces the interaction cost of repeated searches and clicks by providing precise, decision-oriented support. CogSearch has since been fully deployed on the JD platform.

\section{Conclusion}

we presented CogSearch, a system that successfully transforms e-commerce search into a proactive and cognitively aligned decision-making environment. Through our multi-agent framework—which integrates intent decomposition, cross-source information fusion, and interactive steering—we effectively minimize cognitive friction and simplify the consumer journey. Our large-scale deployment on JD.com yields irrefutable empirical results: a 5\% decrease in user decision costs and a statistically significant lift in average UCVR by 0.41\%, soaring to a 30\% increase for complex,  decision-heavy queries. Such performance underscores both the technical efficacy and the commercial potential of our approach. 

Future work will prioritize evolving CogSearch into a lifelong, multimodal shopping companion by integrating sensory perception and long-term cognitive memory, while simultaneously optimizing the computational efficiency of multi-agent reasoning without compromising depth.

\section*{Presenter Biography}
\label{bio}
\textbf{Zhouwei Zhai} is a Scientist at JD.com, focusing on LLM-powered search systems and AI Agents. At JD.com, he spearheaded the transition towards LLM-augmented e-commerce search and led the end-to-end construction of the platform's next-generation AI Search Assistant.
\bibliographystyle{ACM-Reference-Format}
\bibliography{sample-base}

\end{document}